\documentclass[12pt]{article}

\usepackage{epsf}
\usepackage{latexsym,euscript}
\usepackage{amsmath}
\usepackage{amsfonts}
\usepackage{amssymb, epsfig}

\usepackage{color,soul}

\begin{document}

\begin{center}
{\Large \textbf{The Polyakov loop models in the large $N$ limit: \\ 
Phase diagram at finite density}}

\vspace*{0.6cm}
\textbf{O.~Borisenko${}^{\rm a}$\footnote{email: oleg@bitp.kiev.ua},
V.~Chelnokov${}^{\rm a,b}$\footnote{email: chelnokov@itp.uni-frankfurt.de},
S.~Voloshyn${}^{\rm a}$\footnote{email: billy.sunburn@gmail.com}}

\vspace*{0.3cm}
{\large \textit{${}^{\rm a}$ Bogolyubov Institute for Theoretical
Physics, National Academy of Sciences of Ukraine, 03143 Kyiv, Ukraine}} \\
{\large \textit{${}^{\rm b}$ Institut f\"ur Theoretische Physik, Goethe-Universit\"at
Frankfurt, 60438 Frankfurt am Main, Germany}}
\end{center}

\begin{abstract}
The 't Hooft-Veneziano limit of various $U(N)$ and $SU(N)$ Polyakov loop models 
is constructed at finite temperature and non-zero baryon chemical potential. 
In this paper we calculate the free energy, its derivatives, the Polyakov loop expectation values and the baryon density. The phase diagram is described in details.
\end{abstract}

\section{Introduction}

In this and subsequent paper we study the lattice models whose partition function can be written as
\begin{equation}
Z_G(N,N_f) \ = \ \int_G \ \prod_x dU(x) \ \exp \left [ S \left ( {\rm Tr} U(x), {\rm Tr} U^{\dagger}(x)  \right )   \right  ] \ ,  
\label{Gint_def}
\end{equation} 
where $U \in G=U(N), SU(N)$. $\mbox{Tr} U$ will denote the character of the fundamental representation of $G$ and $dU$ - the normalized Haar measure on $G$. 
$N$ is a number of colors and $N_f$ is a number of quark flavors.
The action $S$ can take an arbitrary form including a non-local interaction and any powers of traces. The only restriction is that the action depends on the group elements $U$ in the fundamental and/or adjoint representation.
We present a solution of such models in the 't Hooft-Veneziano limit 
\cite{Hooft_74,Veneziano_76}:
$g\to 0, N\to\infty, N_f\to\infty$ such that the product $g^2 N$ and the ratio 
$N_f/N=\kappa$ are kept fixed ($g$ is the coupling constant).  

Models of type (\ref{Gint_def}) can be regarded as effective Polyakov loop models describing  lattice gauge theory (LGT) at finite temperatures and non-zero chemical 
potentials. Such effective models constitute an important tool in the studies of QCD phase diagram in a certain range of parameters, see \cite{philipsen_rev_19} for a review. In particular, for many such models one can construct exact dual representations \cite{Gattringer11,un_dual18,pl_dual20}. Such dual forms are free of the sign problem and can be reliably studied via Monte-Carlo simulations \cite{Philipsen12,mcdual_21}. Furthermore, these models can be used to investigate different aspects of the large $N$ behavior of finite-temperature LGTs \cite{damgaard_patkos,christensen12,pisarski12,pisarski18,philipsen_quarkyon_19,largeN_sun}.    

The general action we deal with here is written down as  
\begin{eqnarray}
	\label{PL_action}
S &=& \sum_{x,y} S_g(x,y) +
N_f \frac{h}{2} \sum_x \left ( e^{\mu} {\rm {Tr}}U(x) +  e^{-\mu} {\rm {Tr}}U^{\dagger}(x) \right ) \ , \\ 
S_g(x,y) &=& \sum_{n=1} \ {\rm Re} \ \left ({\rm Tr} U(x) \right )^n  K_n(x-y) 
\left ({\rm Tr} U^{\dagger}(y) \right )^n \ . 
\end{eqnarray}
It describes the interaction between Polyakov loops ${\rm {Tr}}U(x)$ at finite temperature and in the presence of $N_f$ heavy degenerate quark flavors. Parameter $h$ can be related to the quark mass (its exact form depends of the kind of lattice fermions) and $\mu=\beta\mu_q$, where $\mu_q$ is the quark chemical potential. 
A well-known simplest case corresponds to the choice 
\begin{equation}
K_n(x-y)=\delta_{n,1} \sum_{\nu=1}^d\delta_{y,x+e_{\nu}} \beta_{eff} \ , 
\label{local_action} 
\end{equation}
where effective coupling depends on the 't Hooft coupling constant $g^2 N$ and a lattice extent in the temporal direction.  An example of the model with a non-local interaction between Polyakov loops has been derived in \cite{greensite16}
(and Refs. therein) via the relative weight method. In all cases
the full action of the model involves only fundamental characters. 

This and other similar effective actions have been a subject
of numerous analytic and numerical investigations. In particular, the large $N$ limit 
with the action (\ref{local_action}) has been studied in Ref.\cite{damgaard_patkos} at zero $\mu$ and in  Ref.\cite{christensen12} at finite $\mu$ using the mean-field approach which is believed to be exact in this limit due to the  factorization property. The resulting mean-field integrals have been computed over $U(N)$ group. 
In Ref.\cite{largeN_sun} we have shown that when $\mu$ is non-vanishing the $SU(N)$ integrals essentially differ from $U(N)$ ones. Therefore, the analysis of \cite{christensen12} should be re-examined for $SU(N)$ model, and this is a main goal of the present paper. 

The phase structure of $U(N)$ model coincides in the large $N$ limit with the one of $SU(N)$ model if $\mu=0$. It can be briefly summarized as follows. 
When $h=0$ (infinitely heavy quarks) one finds a 1st order phase transition. This holds for all $N\geq 3$. If $S_g(x,y)=0$ the model reduces to a two-dimensional LGT in the thermodynamic limit. It exhibits a third order phase transition of the Gross-Witten-Wadia (GWW) type \cite{gross_witten,wadia}. Combined system also possesses GWW type of phase transition \cite{damgaard_patkos}.  
More general local effective actions at zero chemical potential have been introduced 
and solved in Refs.\cite{pisarski12,pisarski18}. These actions include Polykov loops 
in higher representations. 

This paper is organized as follows. In Sec.~2 we derive a large $N$ representation for the partition and correlation functions of Polyakov loop models. 
In Sec.~3 the free energy and other local observables are calculated and the phase structure is described in details. Summary is presented in Sec.~4.

\section{Large $N$ representation}

In this Section we derive a representation for the partition and correlation functions valid
in the large $N$ limit. This representation will serve as a starting point for detailed study
of the critical behavior.

\subsection{Jacobian}

Consider the following change of variables in the partition
function (\ref{Gint_def}):
\begin{eqnarray}
\label{real_U}
\rho(x) \cos\omega(x) \ = \ \frac{1}{N} \ \mbox{Re} \, \mbox{Tr} U(x) \ ,  \\
\rho(x) \sin\omega(x) \ = \ \frac{1}{N} \ \mbox{Im} \, \mbox{Tr} U(x) \ .
\label{im_U}
\end{eqnarray}
The partition function (\ref{Gint_def}) can then be rewritten in the presence of sources as
\begin{eqnarray}
\label{PF_sun_eff}
&&Z_G(N, N_f) =  \int_0^1 \prod_x \rho(x) d\rho(x)  \ \int_0^{2\pi}
\prod_x d\omega(x) \  \prod_x \rho(x)^{\eta(x)+\bar{\eta}(x)} \ e^{i\omega(x)(\eta(x)-\bar{\eta}(x))} \nonumber \\
&&\times \exp \left [ S \left ( N \rho(x) e^{i\omega(x)}, N \rho(x) e^{-i\omega(x)}  \right )   \right  ] \
\prod_x \Sigma (\rho(x),\omega(x)) \ ,
\end{eqnarray}
where $\Sigma(\rho,\omega)$ is the Jacobian of the transformation given by
\begin{eqnarray}
\Sigma(\rho,\omega)  =  N^2 \int_G  dU  \
\delta \left (N\rho \cos\omega -  \mbox{Re} \, \mbox{Tr} U \right ) \
\delta \left (N\rho \sin\omega -  \mbox{Im} \, \mbox{Tr} U \right ) \ .
\label{Jacobian}
\end{eqnarray}
Using the integral representation for the deltas in the last expression,
one can perform the invariant integration with the help of methods of Ref.\cite{largeN_sun}. We find
\begin{equation}
\Sigma(\rho,\omega)  \ = \  \frac{4 N^4}{2\pi} \sum_{q=-\infty}^{\infty} \ e^{-iqN\omega} \ N^{|q|N} A_N(|q|)
\ \Phi_q(\rho) \ ,
\label{Jst}
\end{equation}
where
\begin{equation}
\Phi_q(\rho)  \ = \ \int_{-\infty}^{\infty} e^{-2iN^2\rho s +|q|N \ln (t+is)} \
H_N\left ( 2i\sqrt{t^2 + s^2}, |q| \right ) \  dt ds \ ,
\label{Jnst}
\end{equation}
\begin{equation}
A_N(q) \ = \ \frac{G(N+1)G(q+1)}{G(N+q+1)} \ ,
\label{ANq_def}
\end{equation}
\begin{equation}
H_N\left ( 2ir, |q| \right ) = \exp \left [ -N^2 \left ( r^2 + P(|u|,r)  \right )  \right ] \ , \ u=\frac{q}{N} \ .
\label{HN_def}
\end{equation}
Here, $G(n)$ is the Barnes function and $P(u,r)$ is given by
\begin{equation}
P(u,r) = \sum_{k=0}^{\infty} \ r^{2k+2} \ C_k(u) \ ,
\label{Pxr}
\end{equation}
\begin{equation}
C_k(u) = \frac{(-4)^k}{k+1} \ \sum_{m=1}^{\infty} \ (-u)^m  \
\frac{\Gamma[\frac{m}{2} + k +1] \Gamma[m + 2 k]}{
\Gamma[\frac{m}{2}+1] \Gamma[m] \Gamma[k+2] \Gamma[2 k+2]} \ .
\label{Ck}
\end{equation}
Large $N$ expansion of the function $A_N(q)$ can be calculated by making use of 
the well-known asymptotic expansion for the Barnes function. One gets for the leading term
\begin{equation}
N^{q N} \ A_N(q) \ = \ e^{N^2 f(u) + {\cal{O}}(N)} \ , \ u=q/N \ ,
\label{ANq_as}
\end{equation}
\begin{equation}
f(u) \ = \ \frac{3}{2} \ u - \frac{1}{2} \ (1+u)^2\ln (1+u) + \frac{1}{2} \ u^2 \ln u  \ .
\label{fx_as}
\end{equation}
Combining all results together and replacing summation over $q$ by integration in the large $N$ limit we obtain
\begin{equation}
\Sigma(\rho,\omega)  \ = \ {\rm {const}} \ \int_{-\infty}^{\infty} du dt ds \
e^{N^2 V(\rho,\omega;u,t,s)} \ ,
\label{Sigma_res}
\end{equation}
where
\begin{eqnarray}
V(\rho,\omega;u,t,s) &=& -iu\omega + f(|u|) -2i\rho s + |u|\ln (t+is)  \nonumber  \\
&-& t^2 -s^2 - \sum_{k=0}^{\infty} (t^2+s^2)^{k+1} C_k(|u|)  \ .
\label{Vpot}
\end{eqnarray}
For the $U(N)$ model the last equation simplifies to
\begin{eqnarray}
\frac{1}{N^2} \ln \Sigma(\rho,\omega) \ = \  V(\rho) \ = \
\begin{cases}
- \rho^2   \ , \    \ \ \ \rho \leq \frac{1}{2}  \ ,   \\
\frac{1}{2} \ \ln (1-\rho) - \frac{1}{4} + \frac{1}{2} \ln 2  \ , \ \ \ \rho\geq \frac{1}{2} \ .
\end{cases}
\label{Vr_2dun}
\end{eqnarray}

\subsection{Action}

We assume the Polyakov loop model depends only on the fundamental characters
of $SU(N)$. Therefore, the part of the action arising from the pure gauge sector is
\begin{equation}
S_g = N^2 \ S_g(\{ \rho(x) e^{i\omega(x)}, \rho(x) e^{-i\omega(x)} \} ) \ .
\label{action_gauge}
\end{equation}
Moreover, due to $Z(N)$ and charge conjugation symmetries the action depends only on
the difference $\omega(x)-\omega(y)$. Also, we have assumed that $S_g$ scales at most as $N^2$ at large $N$. This is the case for all conventional lattice models.
The contribution of $N_f$ degenerate quark flavors is taken as in (\ref{PL_action}).
In the 't Hooft-Veneziano limit the ratio $N_f/N=\kappa$ is kept fixed and the quark contribution becomes
\begin{equation}
S_q = N^2 \alpha \sum_x \rho(x) \cos (\omega(x) - i\mu) \ , \ \alpha = \kappa h \ .
\label{action_quark1}
\end{equation}

\subsection{Partition and correlation functions}

Collecting formulas for the action and for the Jacobian from previous subsections we write down the correlation function in the form
\begin{eqnarray}
\Gamma(\eta(x),\bar{\eta}(x)) &=& \left \langle \prod_x \left ( \frac{1}{N}\mbox{Tr} U(x) \right )^{\eta(x)} \  \left ( \frac{1}{N}\mbox{Tr} U^{\dagger}(x) \right )^{\bar{\eta}(x)} \right \rangle  \nonumber   \\
 &=&  e^{\mu \sum_x(\bar{\eta}(x) - \eta(x))} \ \left \langle \prod_x \rho(x)^{\eta(x)+\bar{\eta}(x)} \ e^{i\omega(x)(\eta(x)-\bar{\eta}(x))}  \right \rangle
 \  .
\label{corr_1}
\end{eqnarray}
We performed the global shift $\omega(x) \to \omega(x) + i\mu$. Expectation value in the last expression refers to the following partition function
\begin{equation}
Z = \prod_x \ \int_0^1\rho(x)d\rho(x) \int_0^{2\pi} \frac{d\omega(x)}{2\pi}
\int_{-\infty}^{\infty} \ du(x) dt(x) ds(x) \ e^{N^2 S_{eff}} \ .
\label{PF_largeN}
\end{equation}
The effective action is given by
\begin{eqnarray}
S_{eff} = S_g(\{ \rho(x) e^{i\omega(x)}, \rho(x) e^{-i\omega(x)} \} ) +
\alpha \ \sum_x \rho(x) \cos\omega(x) + \mu \sum_x u(x) \nonumber   \\
+ \sum_x \  V(\rho(x),\omega(x);u(x),t(x),s(x))  \ .
\label{Seff_1}
\end{eqnarray}

\section{Large $N$ solution}

It is now straightforward to get the large $N$ solution. One has to look for translationally invariant saddle-points of all integrals in (\ref{PF_largeN}) that  provide maximum of the effective action. On constant configurations $S_{eff}$ reads
\begin{eqnarray}
S_{eff} = S_g(\rho) + \alpha  \rho \cos\omega + \mu u
-iu\omega + f(|u|) -2i\rho s   \nonumber   \\
+ |u|\ln (t+is) - t^2 -s^2 - \sum_{k=0}^{\infty} (t^2+s^2)^{k+1} C_k(|u|)  \ .
\label{Seff_2}
\end{eqnarray}
Coefficients $C_k(|u|)$ are given in Eq.(\ref{Ck}) and the function $f(u)$ in (\ref{fx_as}). The system of the saddle-point equations is given by
\begin{eqnarray}
\label{se1}
\alpha \rho \sin\omega &=& - i u  \ , \\
\label{se2}
S^{(1)}(\rho) + \alpha \cos\omega &=& 2 i s  \ ,  \\
\label{se3}
\frac{u}{t+is} &=& 2 t E(t,s,u)  \ ,  \\
\label{se4}
2 \rho - \frac{u}{t+is} &=& 2 i s E(t,s,u)  \ ,  \\
\label{se5}
\mu + \frac{\partial f(u)}{\partial u} -i \omega + \ln (t+is)
&=& \sum_{k=0}^{\infty} (t^2+s^2)^{k+1} \frac{\partial C_k(|u|)}{\partial u}  \ ,
\end{eqnarray}
where $S^{(k)}(\rho)=\frac{\partial^k S_g(\rho)}{\partial \rho^k}$ and
\begin{equation}
E(t,s,u) = 1 + \sum_{k=0}^{\infty} (t^2+s^2)^k (k+1) C_k(|u|)  \ .
\label{E_def}
\end{equation}
For the $U(N)$ model the effective action and saddle-point equations take the form
\begin{eqnarray}
S_{eff} = S_g(\rho) + \alpha  \rho \cos\omega + V(\rho)  \ ,
\label{Seff_UN}
\end{eqnarray}
\begin{eqnarray}
\label{se1_un}
\alpha \rho \sin\omega &=& 0  \ , \\
\label{se2_un}
S^{(1)}(\rho) + \alpha \cos\omega  + \frac{\partial V(\rho)}{\partial \rho} &=& 0 \ .
\end{eqnarray}
The function $V(\rho)$ is given in Eq.(\ref{Vr_2dun}).
Let us stress the systems of equations are valid for any choice of the gauge action $S_g$
both for $U(N)$ and $SU(N)$ Polyakov loop models.
The simplest action is of the form $S_g = d\beta \rho^2$.

\subsection{Free energy and phase structure of $U(N)$ model}

First, we treat the simpler example of $U(N)$ model and start with two remarks
\begin{itemize}
\item
When $S_g(\rho)=0$ the model reduces to the two-dimensional LGT and Eqs.(\ref{se1_un}),
(\ref{se2_un}) describe the one-plaquette integral in the large $N$ limit. Indeed, one
can easily recover the GWW solution of this model by solving these equations.

\item
When $S_g = d\beta \rho^2$ we recover the solution of Refs.\cite{damgaard_patkos} and \cite{christensen12}. When $\alpha=0$ one finds the first order phase transition which turns into the third order when $\alpha$ is nonzero. 

\end{itemize}
In the general case one has from (\ref{se1_un}) $\omega=0$. Using  (\ref{Vr_2dun}),  Eq.(\ref{se2_un}) becomes 
\begin{eqnarray}
S^{(1)}(\rho) + \alpha \ = \
\begin{cases}
2 \rho   \ , \    \ \ \ \rho \leq \frac{1}{2}  \ ,   \\
\frac{1}{2(1-\rho)} \ \ , \ \ \ \rho\geq \frac{1}{2} \ .
\end{cases}
\label{Eq_gen_un}
\end{eqnarray}
Note that while the potential $V(\rho)$ and its first two derivatives are continuous
in the point $\rho=1/2$, the third derivative exhibits a finite jump $\Delta=8$. It might be an evidence that, independently of the interaction in the original action, the large $N$ limit always possesses a third order phase transition.
This is the case, indeed.
Expanding the effective action and the saddle-point equation around point $\rho=1/2$
one finds for the solution
\begin{equation}
\rho_s \ \approx \ \frac{1}{2} + \frac{S^{(1)}\left ( \frac{1}{2} \right ) + \alpha - 1}{2-S^{(2)}\left ( \frac{1}{2} \right ) } \ .
\label{rho_gen}
\end{equation}
and for the effective action which becomes the free energy
\begin{eqnarray}
S_{eff} &=& S_g(\rho_s) + \frac{\alpha}{2} - \frac{1}{4} + \frac{1}{2} \ \frac{\left ( S^{(1)}\left ( \frac{1}{2} \right ) + \alpha - 1 \right )^2}{2-S^{(2)}\left ( \frac{1}{2} \right )}   \nonumber  \\
&+& \frac{1}{6} \ \left ( \frac{ S^{(1)}\left ( \frac{1}{2} \right ) + \alpha - 1 }{2-S^{(2)}\left ( \frac{1}{2} \right )} \right )^3 \ \left ( S^{(3)}\left ( \frac{1}{2} \right ) + V^{(3)}\left ( \frac{1}{2} \right )  \right ) \ ,
\label{Seff_UN_solution}
\end{eqnarray}
where $V^{(3)}\left ( \frac{1}{2}-0 \right )=0$ and  $V^{(3)}\left ( \frac{1}{2}+0 \right )=-8$.
The critical surface in the space of all possible couplings turns out to be
\begin{equation}
S^{(1)}\left ( \frac{1}{2} \right ) + \alpha \ = \ 1 \ .
\label{crit_surf_un}
\end{equation}
First two derivatives of the free energy are continuous, as expected. The third one exhibits a jump $\Delta = 8$. Therefore, the system always undergoes a 3rd order phase transition. This transition, however is not unique. In particular, one encounters a 1st order phase transition when $\alpha$ is sufficiently small. 
A border between 3rd and 1st order transitions can be determined by solving the following system
\begin{eqnarray}
	\begin{cases}
		S^{(1)}\left( \frac{1}{2} \right ) + \alpha - 1 =0  \ ,  \\
		2 - S^{(2)}\left( \frac{1}{2} \right ) = 0 \ . 
		\label{crit_eqsys_un}
	\end{cases}
\end{eqnarray}
Let us consider few simple examples with ferromagnetic couplings. 
If $S_g(\rho)$ is quadratic in $\rho$, {\it i.e.} $S = b \rho^2$ we have
\begin{eqnarray}
	1  - b - \alpha =0 \ , \  1 - b  = 0 \ . 
\label{un_example1}
\end{eqnarray} 
Here, $b=d \beta_{eff}$ for the local action (\ref{local_action}) and $b=K(0)$ for the non-local action of Ref.\cite{greensite16}, where $K(0)$ is the zeroth mode of the kernel $K(x-y)$. The solution $b = 1, \alpha = 0$ describes a point on the critical line 
$1  - b - \alpha =0$ where the system undergoes the 1st order transition. 
This result is in full agreement with Refs.\cite{damgaard_patkos,christensen12}. 

If $S_g(\rho) = b_1 \rho^2+ b_2 \rho^4$ one gets from (\ref{crit_eqsys_un})
\begin{eqnarray}
2 - 2b_1 - b_2 - 2 \alpha  = 0 \ , \ 2 - 2 b_1 - 3 b_2 = 0 \ . 
\label{un_example2}
\end{eqnarray}
The solution of this system $b_1 = \frac{1}{2} (2 - 3 \alpha), b_2 = \alpha$
gives a line (intersection of two  planes) that belongs to the critical surface 
$2 - 2 b_1- b_2 - 2 \alpha=0$ where the 3rd order transition appears. 
Intersection of the plane $b_2=\alpha$ and the critical surface one finds a 2nd order phase transition. Model with such action has been studied before in \cite{pisarski18}. Our results fully agree with this paper. 

In general, for $k$ ferromagnetic couplings $b_i, i=1,\ldots,k$ 
the solution of the system (\ref{crit_eqsys_un}) defines a $k$-dimensional plane that lies on the $(k+1)$-dimensional critical surface. This plane is a border between regions of 3rd and 1st order phase transitions where the system 
undergoes a 2nd order transition.

\subsection{Free energy and phase structure of $SU(N)$ model}

Here we turn our attention to the $SU(N)$ model and analyze the system of equations
(\ref{se1})-(\ref{se5}). In what follows we assume $\mu\geq 0$.
From (\ref{se1}), (\ref{se2}) and an obvious combination of
(\ref{se3}) and (\ref{se4}) we obtain solutions on $\omega$, $s$ and $t$ variables
\begin{eqnarray}
\label{sol_omega}
\cos\omega &=&  \sqrt{1+\frac{u^2}{\alpha^2 \rho^2}}  \ , \\
\label{sol_s}
i s &=& \frac{1}{2} \ S^{(1)}\left ( \rho \right ) + \frac{1}{2 \rho} \
\sqrt{u^2 + \alpha^2 \rho^2} \ , \\
\label{sol_t}
t &=& \frac{u}{2 \rho} \ .
\end{eqnarray}
Eqs.(\ref{se3}) and (\ref{se5}) are used to construct solutions for $\rho$ and $u$
variables, correspondingly. As will be seen right below, to reveal the critical behavior
it is sufficient to study these equations in the vicinity of the point $u=0$.
Expanding the right-hand side of (\ref{se3}) in powers of $u$ we write down the solution for $\rho$ as
\begin{equation}
\rho \ = \ \rho_0 + \rho_1 u  + {\cal {O}}\left ( u^2  \right ) \ .
\label{sol_rho}
\end{equation}
One finds
\begin{equation}
\rho_1 \ = \ \frac{1}{\rho_0} \ \frac{\sqrt{1-4\rho_0^2}}{2-S^{(2)}\left ( \rho_0 \right )} \ ,
\label{sol_rho_1}
\end{equation}
$\rho_0$ satisfies the following equation
\begin{equation}
S^{(1)}\left ( \rho_0 \right ) + \alpha - 2 \rho_0 \ = \ 0 \ .
\label{eq_rho_0}
\end{equation}
Substituting this solution into (\ref{se5}) we get the following equation on $u$ up to
${\cal {O}}\left ( u^2  \right )$
\begin{equation}
z + u \ln u - b_1 u  = 0 \ ,
\label{eq_u}
\end{equation}
where
\begin{equation}
z \ = \ \mu + \sqrt{1-4 \rho_0^2}  + \ln 2 \rho_0 -
\ln\left ( 1 +  \sqrt{1-4 \rho_0^2}  \right ) \ ,
\label{z_def}
\end{equation}
\begin{equation}
b_1 \ = \  -1 + \frac{1}{\alpha \rho_0} + \frac{1}{2} \ \ln (1-4\rho_0^2) -
\frac{1-4\rho_0^2}{\rho_0^2 \left ( 2 - S^{(2)}\left ( \rho_0 \right )  \right )} \ .
\label{b1_def}
\end{equation}
Now we turn to Eq.(\ref{eq_u}).
It follows from (\ref{z_def}) that if $\rho\leq 1/2$ and $\mu$ is sufficiently small there is no solution to this equation,
maximum of the integrand is achieved for $u=0$ and the free energy equals to the $U(N)$ free energy. When $\mu$ grows a non-trivial solution appears for $z\geq 0$
which provides maximum of the effective action. Therefore, the approximate solution
in the vicinity of the critical line can be expanded in a power series of $z$.

With notation
\begin{equation}
y \ = \ - \frac{z}{W_{-1} \left ( - e^{-b_1} z \right ) } \  \ , \   \
\label{notation_sun1}
\end{equation}
the solution can be written as
\begin{eqnarray}
u  =  -y - \frac{b_2}{1 + W_{-1} \left ( - e^{-b_1} z \right )}\ y^2
+ {\cal O}\left ( y^3 \right ) \ ,
\label{u_solut_small_h}
\end{eqnarray}
where $W_{-1}(x)$ is the lower branch of the Lambert function. 
$b_2$ is a function of $\rho_0$ and $\alpha$ exact form of which is not important here. Combining all solutions together and substituting them into the effective action
we obtain for the free energy the following simple result
\begin{eqnarray}
F = (\alpha - \rho_0) \rho_0 + S(\rho_0) -
\frac{1}{4} \  \left ( 1 + 2 W_{-1} \left ( - e^{-b_1} z \right ) \right ) \ y^2
+ {\cal O}\left ( y^3 \right ) .
\label{Fren_sun_small_h}
\end{eqnarray}
One finds a third order phase transition along the critical line which is given by the equation
\begin{equation}
z \ = \ 0 \  \  \Rightarrow \  \mu \ = \  \ln \frac{1 + \sqrt{ 1 - 4 \rho_0^2}}{2 \rho_0} - \sqrt{1 - 4 \rho_0^2} \ .
\label{critical_line}
\end{equation}

\begin{figure}[htb]
	\centerline{\epsfxsize=7cm \epsfbox{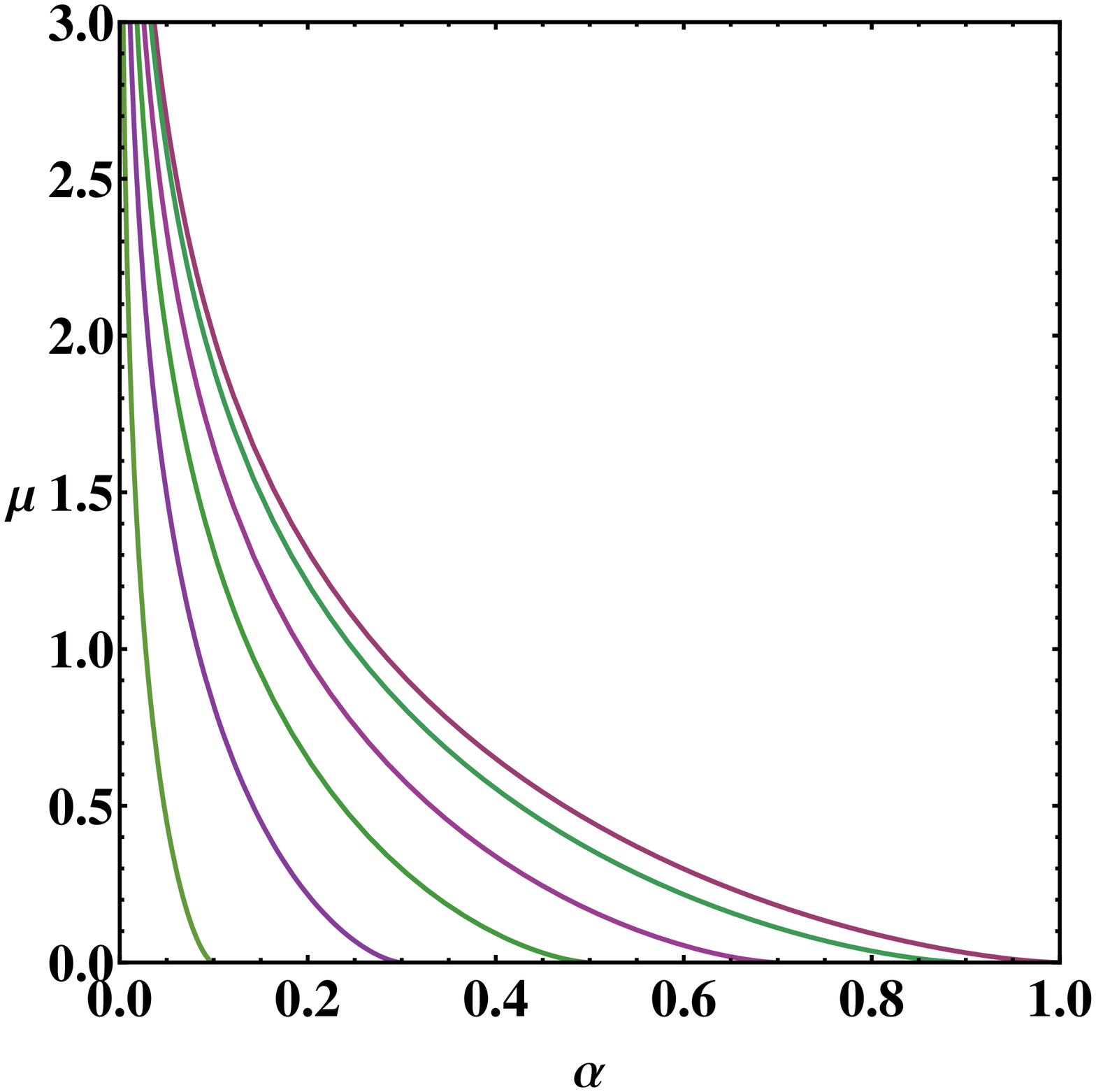} \epsfxsize=7cm \epsfbox{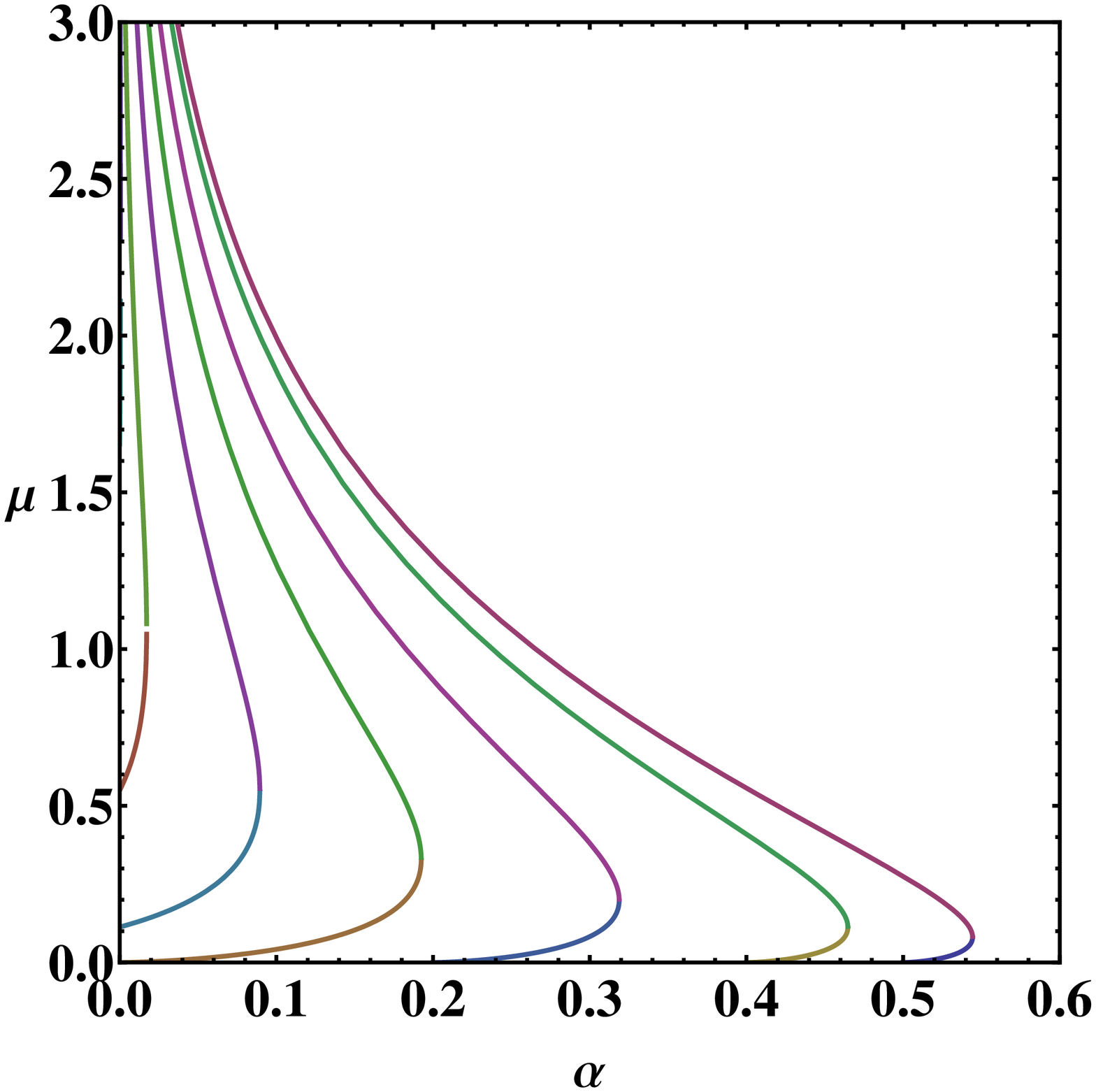} }
	\caption{\label{sun_crit_line} Plots of critical line (\ref{critical_line}) in the coordinates $\alpha-\mu$ and fixed $b$. 
		Left panel: action $S =b\rho^2$. 
		Right panel: action $S = b \rho^2+ b_2 \rho^4$ for $b_2=1$. 
		On both panels lines from right to left are $b=0.1, 0.3, 0.5, 0.7, 0.9, 0.99$.}
\end{figure}

Eq.(\ref{eq_rho_0}) and subsequent solutions allow us to analyze the $SU(N)$ model with an arbitrary action $S_g(\rho)$. 
Consider the simplest example with $S_g(\rho)= b \rho^2$. One finds $\rho_0= \frac{\alpha}{2(1- b)}$. Third derivative of the free energy (\ref{Fren_sun_small_h}) exhibits an infinite jump on the surface described by (\ref{critical_line}). 
In the limit $\mu \to 0$ we enter $U(N)$ regime. Here the third derivative of the free energy exhibits a finite jump as described in previous subsection. 
The critical line is shown on the left panel of Fig.\ref{sun_crit_line} 
for various values of $b$. When $b$ decreases the line tends to the critical line of the one-site model \cite{largeN_sun}. 

The model with the action $S_g(\rho) = b_1 \rho^2 + b_2 \rho^4$ has three solutions
\begin{eqnarray}
	\rho_0 \ = \  \begin{cases}
		-\frac{1}{\sqrt[3]{p}}(b_1-1)  + \frac{1}{6 b_2} \sqrt[3]{p} \ , \\
		\frac{\left(1 \pm i \sqrt{3}\right) }{2 \sqrt[3]{p}}(b_1-1)-\frac{\left(1 \mp i \sqrt{3}\right) }{12 b_2}\sqrt[3]{p} \ ,
		\label{crit_eq syytyun}
	\end{cases}
\end{eqnarray}
where $p=3 \left(\sqrt{3 b_2^3 \left(27 \alpha^2 b_2+8 (b_1-1)^3\right)}-9 \alpha b_2^2\right)$. It can be studied similarly to the previous example. 
The critical line in this case is shown on the right panel of Fig.\ref{sun_crit_line} 
for various values of $b_1$. The different colors of the curve parts seen for smaller values of $\mu$ correspond to different branches in the solution (\ref{crit_eq syytyun}) for $\rho_0$. On the lower branch of the curve one finds a first order phase transition. This branch joins smoothly to the upper branch where the third order transition occurs. Two  regions with different type of phase transitions are 
separated by the surface defined as  
\begin{equation} 
b_2^3 \left [ 27 \alpha^2 b_2 + 8 (b_1- 1)^3 \right ] =0 \ . 
\label{border_line_r4}
\end{equation}

\begin{figure}[htb]
	\centerline{\epsfxsize=7cm \epsfbox{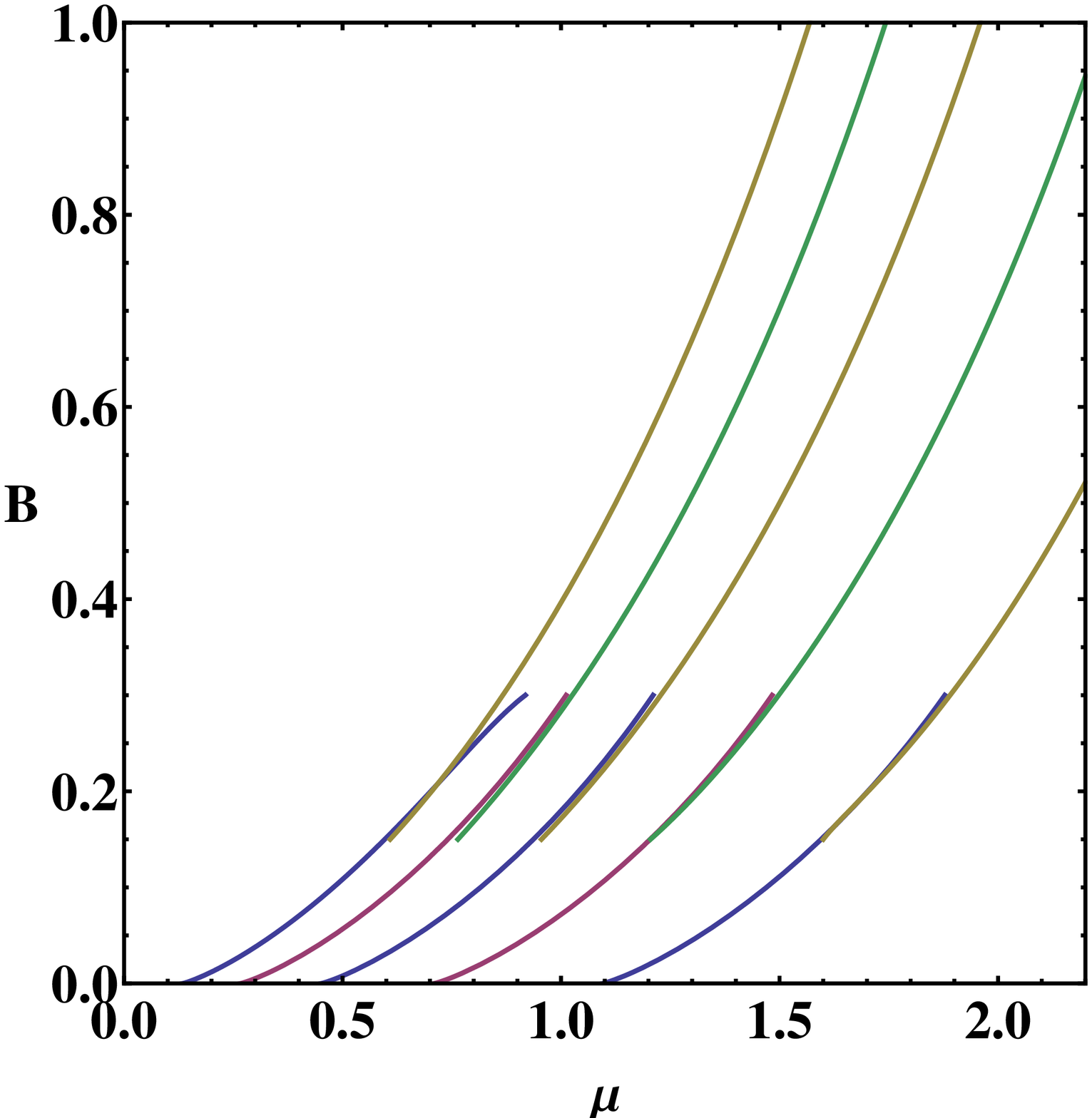} \epsfxsize=7cm \epsfbox{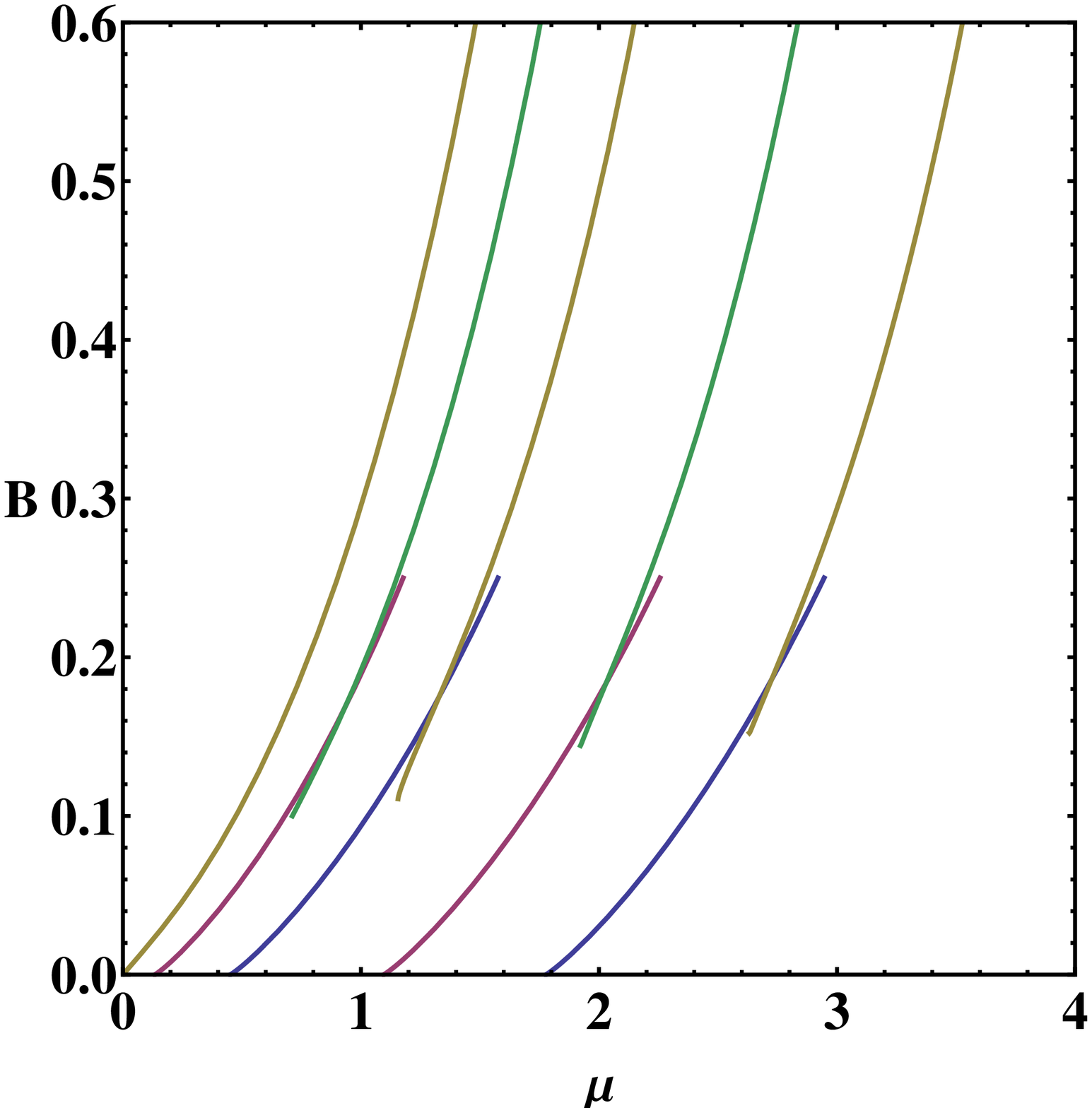} }
	\caption{\label{sun_density} Plots of particle density $B$ as a function of  $\mu$ for the action $S =b\rho^2$. Left panel: $b=0.2$; from left to right $\alpha=0.6, 0.5, 0.4, 0.3, 0.2$. Right panel: $b=0.6$; from left to right $\alpha=0.4, 0.3, 0.2, 0.1, 0.05$.
		}
\end{figure}

In the region $\rho_0>1/2$, corresponding to large values of ferromagnetic couplings in $S_g(\rho)$ and/or large values of $\alpha$, the solution can be found with the help of the following expansion for the function $P(u,r)$ in (\ref{Pxr}) (see Appendix of Ref.\cite{largeN_sun})
\begin{eqnarray}
	P(u,r) &=& -r^2  + u \left[ 1 - \sqrt{1 - \frac{4 r^2}{u^2}} +
	\ln \frac{1 + \sqrt{ 1 - \frac{4 r^2}{u^2}}}{2}  \right] +  \frac{1}{4} \ln \left ( 1 - \frac{4 r^2}{u^2} \right )  + \nonumber  \\
	&+&  \frac{1}{6 u} \left(\frac{ 1  }{(1 - \frac{4 r^2}{u^2})^{3/2}} -1 \right) +  {\cal O}\left ( u^{-4} \right ) \ .
	\label{Pu_large_series}
\end{eqnarray}
Below we present results only for the standard action $S_g(\rho)=b\rho^2$. 
One obtains the following approximate solutions of Eqs.(\ref{se3}) and (\ref{se5}) with the help of (\ref{Pu_large_series}) 
\begin{eqnarray}
\label{large_b_a_solution}
u &\approx& \rho \alpha \sinh\mu \ , \\ 
\rho &\approx& \frac{2 b- \alpha \cosh \mu +\sqrt{A^2-4 b}}{4 b} \ , \nonumber  
\end{eqnarray}
where $A= 2 b + \alpha \cosh \mu $. It leads to the free energy expansion of the form 
\begin{eqnarray}
F = b + \alpha \cosh \mu  + A\frac{ \sqrt{A^2-4 b}-A}{8 b}  - \frac{1}{2}  -\frac{1}{2} \log \frac{A + \sqrt{A^2-4 b}}{2} + {\cal{O}}\left (A^{-1} \right) \ .
\label{Fren_sun_large_beta}
\end{eqnarray}
Though we could not prove it analytically we think the transition from the regime described by Eq.(\ref{Fren_sun_small_h}) to the regime described by (\ref{Fren_sun_large_beta}) is smooth and is not  accompanied by any phase transition. This is confirmed by our numerical computations of the free energy 
above the critical surface (\ref{critical_line}), 
If so, the large $N$ limit exhibits one third order phase transition
from $U(N)$ to $SU(N)$ regime and whose critical surface is given by Eq.(\ref{critical_line}). Below the critical surface the free energy does not depend 
on the chemical potential $\mu$ and the particle density is vanishing. 
Above the critical surface a non-trivial dependence on $\mu$ appears. The particle density is non-zero and is given by the variable $u$ as 
\begin{equation} 
B = \frac{1}{L^d} \ \frac{\partial \ln Z}{\partial \mu} = u \ . 
\label{part_density}
\end{equation} 
Plots of the quark density as a function of $\mu$ are shown in Fig,\ref{sun_density}. 
The lower branches of each curve are derived from the free energy (\ref{Fren_sun_small_h}) while upper branches - from free energy given by Eq.(\ref{Fren_sun_large_beta}). 

Expectation values of the Polyakov loop and its conjugate are given in the large $N$ limit by $\rho e^{\pm i \omega}$ and can be easily reconstructed from the corresponding solutions for $\rho$ and $\omega$ variables given above.

Finishing this Section we would like to add several comments:
\begin{itemize}
\item
When $S_g(\rho)=0$ the partition function factorizes into a product of one-site integrals. In this case the solutions presented above agree with our solution given 
in \cite{largeN_sun}. When $S_g(\rho)=b \rho^2$ and $\mu=0$ our results coincide with the results of Refs.\cite{damgaard_patkos,christensen12}. 
In particular, the free energy (\ref{Fren_sun_large_beta}) reduces
to the $U(N)$ free energy when $\mu=0$.

\item
We have checked our results using the large $N$ factorization property in the theory with the action $S_g(\rho)= b \rho^2$. 

\item 
We would like to emphasize the universality of third order phase transition: it occurs in all $U(N)$ and $SU(N)$ models independently of the form of interaction between Polyakov loops in $S_g(\rho)$. Such transition is an inherent $U(N)$ group property 
in the large $N$ limit. In the present approach it appears due to large $N$ form of the Jacobian (\ref{Vr_2dun}): its third derivative exhibits a finite jump at the point $\rho=\frac{1}{2}$. So does the free energy when the saddle point reaches the value $\frac{1}{2}$ at some values of the parameters. The importance of the point $\rho=\frac{1}{2}$ in relation to the GWW transition was explained, in a somewhat different mathematical settings, in \cite{pisarski18}.

\item
In the case of an imaginary $\mu$ one can recover again the $U(N)$ solution. Indeed, when $\mu$ is imaginary, there is no
solution for saddle-point equations on the real line for $u$ variable. Therefore, the maximum of the exponent in Eq.(\ref{PF_largeN})
is reached for $u=0$ and this recovers the $U(N)$ case.

\item
In the heavy dense limit $\mu\to\infty, \alpha\to 0$ such that $\alpha e^{\mu}={\rm const}$ one obtains from Eq.(\ref{critical_line})
that the critical point is given by $\rho_0 e^{\mu}=1/e$.
For the standard choice  $S_g(\rho)=b \rho^2$ one finds from Eq.(\ref{eq_rho_0})
$\rho_0=\frac{\alpha}{2(1-b)}$. Therefore, the critical line is described by equation 
$\alpha e^{\mu}=\frac{2}{e} (1-b)$. Moreover, the last relation appears to be universal, in the sense it is valid for all polynomial actions with ferromagnetic couplings. 

\item 
As can be seen from Fig.\ref{sun_density}, the quark density does not show the onset transition, even in the heavy dense limit. This is not surprising because approximation we used for the full static quark determinant is only valid in the region $h\gg e^{\mu}$ ($m\gg \mu$). To see the onset transition and the possible quarkyonic phase one should study the large $N,N_f$ limits of the full static determinant. Such study at large number of colors has been accomplished in 
\cite{philipsen_quarkyon_19}. 

\end{itemize}

\section{Summary}

In this paper we have studied the 't Hooft-Veneziano limit of some Polyakov loop models both at zero and finite chemical potential. Based on our earlier work \cite{largeN_sun} we derived the representation for partition and correlation functions of the model convenient in the large $N$ limit. Using this representation we 
studied phase transitions in $U(N)$ and $SU(N)$ models for very general actions involving fundamental characters of the Polyakov loop. Whenever possible we have compared our results for $U(N)$ models with those available in the literature \cite{damgaard_patkos,christensen12,pisarski12,pisarski18}. 

Our main finding is that at finite density the large $N$ limit of the Polyakov loop models differs for $U(N)$ and $SU(N)$ groups. While the free energy of $U(N)$ model does not depend on the chemical potential the free energy of $SU(N)$ model does, 
even in the large $N$ limit.  
There exists a critical surface in the space of all couplings such that when $\mu$ grows the system passes through this surface. We have found a 3rd order phase transition
of the GWW type across the critical surface. Unlike the genuine GWW phase transition, the third derivative of the free energy is infinite on the critical surface. 
Above the surface the free energy depends on $\mu$, so the non-vanishing quark density appears. However, there is no onset transition which can only exhibit itself 
in the theory with full static quark determinant. 

In the upcoming paper \cite{largeN_sun_II} we use representation for correlation functions of Sec.2 to calculate the two- and $N$-point correlations and the corresponding screening masses. These calculations reveal an interesting physics 
of the region where the free energy depends on $\mu$. 
Other important directions that are currently under investigation are 1) construction of the 't Hooft-Veneziano limit with an exact static quark determinant 
and 2) inclusion Polyakov loops in higher representations of the group.

\vspace{0.5cm}

{\bf \large Acknowledgements}

O. Borisenko acknowledges support from the National 
Academy of Sciences of Ukraine in frames of priority project 
"Fundamental properties of matter in the relativistic collisions 
of nuclei and in the early Universe" (No. 0120U100935). 

The author V. Chelnokov acknowledges support by the Deutsche
Forschungsgemeinschaft (DFG, German Research Foundation) through the CRC-
TR 211 ’Strong-interaction matter under extreme conditions’ – project number
315477589 – TRR 211.

\end{document}